\documentclass[preprint,12pt]{elsarticle}

\usepackage{xcolor}

\usepackage{amssymb}
\usepackage{amsmath}

\journal{Nuclear Physics B}

\begin{document}

\begin{frontmatter}



\title{Thermal stable nonlinear Raman–Nath diffraction and  Cherenkov radiation in  PPKTP crystals} 


\author{Tao Xie}
\affiliation{organization={Key Laboratory of Low-Dimensional Quantum Structures and Quantum Control of Ministry of Education, Department of Physics and Synergetic Innovation Center for Quantum Effects and Applications, Hunan Normal University}, 
            city={Changsha},
            postcode={410081}, 
            state={Hunan},
            country={China}}

\author{Yan-Ming Liu} 
\affiliation{organization={Hubei Key Laboratory of Optical Information and  Pattern Recognition, Wuhan Institute of Technology},
            city={Wuhan},
            postcode={430205}, 
            state={Hubei},
            country={China}}

\author{Wen-Xin Zhu\corref{cor1}} 
\affiliation{organization={College of Precision Instrument and Opto-Electronics Engineering, Key Laboratory of Opto-Electronics Information Technology, Ministry of Education, Tianjin University},
            addressline={}, 
            city={Tianjin},
            postcode={300072}, 
            state={Tianjin},
            country={China}}
\ead{wenxinzhu@tju.edu.cn} 

\author{Xue-Shi Guo\corref{cor1}}
\affiliation{organization={College of Precision Instrument and Opto-Electronics Engineering, Key Laboratory of Opto-Electronics Information Technology, Ministry of Education, Tianjin University}, 
            city={Tianjin},
            postcode={300072}, 
            state={Tianjin},
            country={China}}
\ead{xueshiguo@tju.edu.cn}

\author{Rui-Bo Jin\corref{cor2}} 
\affiliation{organization={Key Laboratory of Low-Dimensional Quantum Structures and Quantum Control of Ministry of Education, Department of Physics and Synergetic Innovation Center for Quantum Effects and Applications, Hunan Normal University},  
            city={Changsha},
            postcode={410081}, 
            state={Hunan},
            country={China}}
\affiliation{organization={Hubei Key Laboratory of Optical Information and  Pattern Recognition, Wuhan Institute of Technology}, 
            city={Wuhan},
            postcode={430205}, 
            state={Hubei},
            country={China}}
\ead{jrb@hunnu.edu.cn}

\cortext[cor1]{Corresponding author (Tianjin University)}
\cortext[cor2]{Corresponding author (Hunan Normal University)}

\begin{abstract}
Nonlinear Raman-Nath diffraction (NRND) and nonlinear Cherenkov radiation (NCR) are significant nonlinear diffraction phenomena in optics. Previous studies have primarily focused on NRND and NCR in uniaxial crystals, particularly in periodically poled lithium niobate (PPLN) crystals. However, research on these phenomena in biaxial crystals, such as periodically poled potassium titanyl phosphate (PPKTP), has been limited, and the study of NCR in PPKTP has not yet been undertaken.
In this work, we experimentally investigated NRND and NCR phenomena in PPKTP crystals under varying incident angles, pump polarizations, poling periods, and crystal temperatures. Our findings indicate that PPKTP exhibits over ten times greater thermal stability compared to PPLN. 
This high thermal stability is promising for applications in parallel optical computing, as it helps reduce optical mode deviations and minimize bit error rates.
 
\end{abstract}

\begin{graphicalabstract}
\end{graphicalabstract}

\begin{highlights}
\item Nonlinear Raman–Nath diffraction (NRND) and nonlinear Cherenkov radiation (NCR)  are experimentally investigated in PPKTP, a biaxial crystal.
\item PPKTP (3 $\mu$m /°C) exhibits over ten times greater thermal stability compared to PPLN (52 $\mu$m /°C). 
\item This high thermal stability is promising for applications in  optical parallel  computing, as it helps reduce optical mode deviations and minimize bit error rates.
\item Biaxial crystals offer substantially richer phase-matching configurations than uniaxial crystals, enabling more versatile NCR and NRND frequency-conversion processes.

\end{highlights}

\begin{keyword}
Nonlinear Raman–Nath diffraction (NRND) \sep nonlinear Cherenkov radiation (NCR) \sep PPKTP



\end{keyword}

\end{frontmatter}




\section{Introduction}
\begin{figure}[!thp]
\centering
\includegraphics[width= 0.99\textwidth]{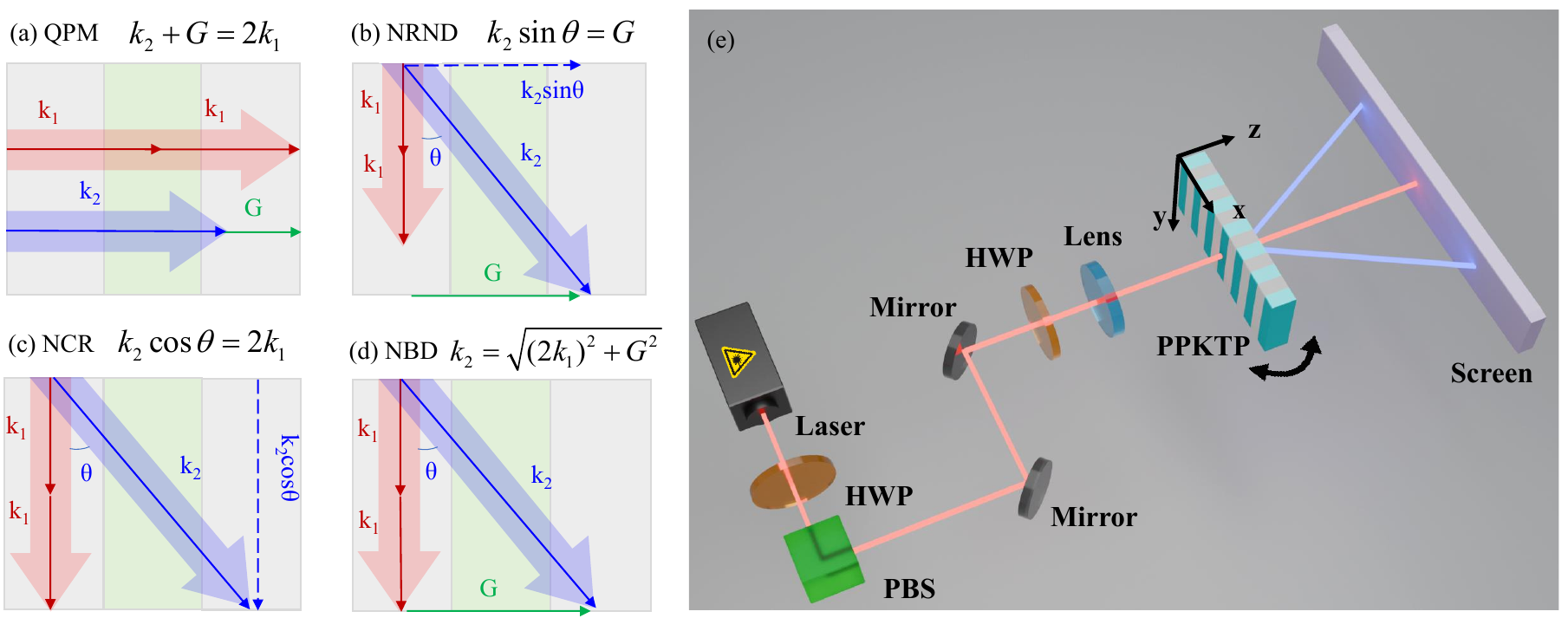}
\caption{Phase-matching conditions and experiment setup. (a) Quasi phase-matching (QPM), (b) Nonlinear Raman–Nath diffraction (NRND), (c) Nonlinear Cherenkov radiation (NCR) and (d) Nonlinear Bragg diffraction (NBD), (e) Experiment setup. Here,  $k_1$ is the wave vector of the pump laser, $k_2$ is the wave vector of the second harmonic laser, $G$ is the reciprocal lattice vector, and $\theta$ is the diffraction angle of the second harmonic.
}
\label{fig:concept}
\end{figure}

%
Second harmonic generation (SHG) is one of the most extensively studied processes in nonlinear optics. To achieve efficient SHG, quasi-phase matching (QPM) condition in periodically poled crystal  is often employed\cite{niu2023COL}.  Besides QPM condition, there are  two different nonlinear process  may occur: nonlinear Raman-Nath diffraction (NRND)\cite{vyunishev2014OL,li2024PhotoniX,shapira2011OL} and nonlinear Cherenkov radiation (NCR)\cite{saltiel2009IEEE,sheng2010OL}.
As shown in Fig.\,\ref{fig:concept} (a-d), NRND arises from transverse phase matching, involving the reciprocal lattice vectors of the periodically poled crystal, and enables multiple noncollinear parametric processes. This phenomenon is analogous to linear grating diffraction, manifesting as a series of diffracted spots. In contrast, NCR is governed by longitudinal phase matching and typically generates only a pair of brighter spots. 
Furthermore, when the diffraction angles of NRND and NCR coincide simultaneously, i.e.,  satisfying both transverse and longitudinal phase-matching conditions, nonlinear Bragg diffraction (NBD) occurs \cite{saltiel2009OL,shapira2011OL}.
NRND and NCR are complementary in nonlinear photonics, collectively broadening the applications of SHG, such as ultrashort pulse characterization\cite{holmgren2007OL,aleksandrovsky2011APL}, high efficiency frequency conversion\cite{ren2013APL}, nondestructive diagnosis of domain structures\cite{deng2010OE,kalinowski2012OL,karpinski2015OE,hong2022PhotonicsResearch}. 
%
%
%
%
%

In the past few decades, significant progress has been made in understanding NRND and NCR phenomena. Previous studies can be roughly classified into two categories. On one hand, many works have investigated how the properties of the pump laser—such as intensity \cite{hong2025UScience}, beam structure \cite{zhang2025COL,buono2022OEA}, beam size \cite{sheng2012JOSAB1}, wavelength \cite{sheng2012JOSAB1,vyunishev2014OL,hong2022PRA}, polarization \cite{sheng2012JOSAB2}, the angle of incidence and azimuthal orientation\cite{saltiel2009OL,zou2025COL,an2013APL} affect NRND and NCR. On the other hand, other studies have focused on how the properties of the nonlinear crystal itself, such as temperature \cite{shapira2011OL}, poling period \cite{li2024PhotoniX}, domain wall thickness \cite{yang2017OE,hong2024JOSAB}, crystal structure \cite{zhao2016OE,saltiel2009IEEE}, and even the crystal dimensions \cite{zhang2021Optica,zhang2018OE}, influence these phenomena.

Although extensive researches has been conducted, previous work has primarily focused on  periodically poled lithium niobate (PPLN) crystals. As a uniaxial crystal, PPLN has a relatively simple refractive index  distribution and phase-matching condition, making theoretical modeling and experimental verification more straightforward. 
In contrast, research on biaxial crystal, such as periodically poled potassium titanyl phosphate (PPKTP) crystal, is still largely unexplored \cite{fragemann2004APL}. Notably, the study of NCR in PPKTP is still lacking, to the best of our knowledge.  This may be partly due to the more complex birefringence properties and phase-matching conditions of PPKTP crystals \cite{Zeng2024COL, Jin2024SCPMA}, as well as the more challenging fabrication processes and integration techniques required.
Compared to PPLN, PPKTP offers a higher optical damage threshold, a broader temperature tolerance bandwidth, and superior resistance to photorefraction, which gives it greater potential for applications involving high-power pumping, fluctuating temperature environments, and frequency conversion processes requiring long-term stability \cite{dmitriev2013handbook}.
Therefore it is very meaningful to investigate the nonlinear diffraction in PPKTP.

In this work, we present a comprehensive study of the NRND and NCR phenomena in PPKTP crystals.
We experimentally used an 810 nm femtosecond pump laser to generate a 405 nm second-harmonic laser in a PPKTP crystal under various conditions, including different incident angles (+10° to $-$10°), pump polarization ($x$- and $y$-polarization), crystal temperatures (ranging from room temperature at 24 °C to 90 °C), and poling periods (9.83 $\mu$m and 3.43 $\mu$m).
Our findings indicate that PPKTP crystals exhibit superior thermal stability compared to PPLN, specifically 
 3 $\mu$m /°C versus 52 $\mu$m /°C. 
This high thermal stability is promising for applications in  optical parallel  computing, as it helps reduce optical mode deviations and minimize bit error rates.

\section{Experiment and results}
We test the NRND and NCR using the experimental setup shown in Fig.\,\ref{fig:concept}\text{(e)} . Femtosecond laser pulses (Mira900, Coherent Co.)  with a repetition rate of 76 MHz, a bandwidth of 5.5 nm and a central wavelength of 810 nm were utilized as the  pump laser.
The pulse passes through a power controller (consisting of an HWP and a PBS) to control the pump power, through two mirrors to collimate the optical path, and through an HWP to control the polarization of the pump (the polarization can be $x$- or $y$-polarization), through a f=50 mm lens to focus the pump  on the PPKTP (laser propagates along the z-axis).
The width of the PPKTP crystal in x, y and z directions is 10 mm, 2 mm, and 1 mm, respectively. The poling period of the  PPKTP crystal is 9.83 $\mu$m. The distance from the PPKTP crystal's surface B to the screen is 54.0 mm.
The PPKTP crystal is placed on a crystal stage that can be rotated and can be temperature controlled. Under different phase matching conditions, a nonlinear diffraction process occurs and the diffracted spot is presented on the screen. We captured the images using a camera (Panasonic Lumix s5).

\begin{figure}[!thp]
\centering
\includegraphics[width= 0.75\textwidth]{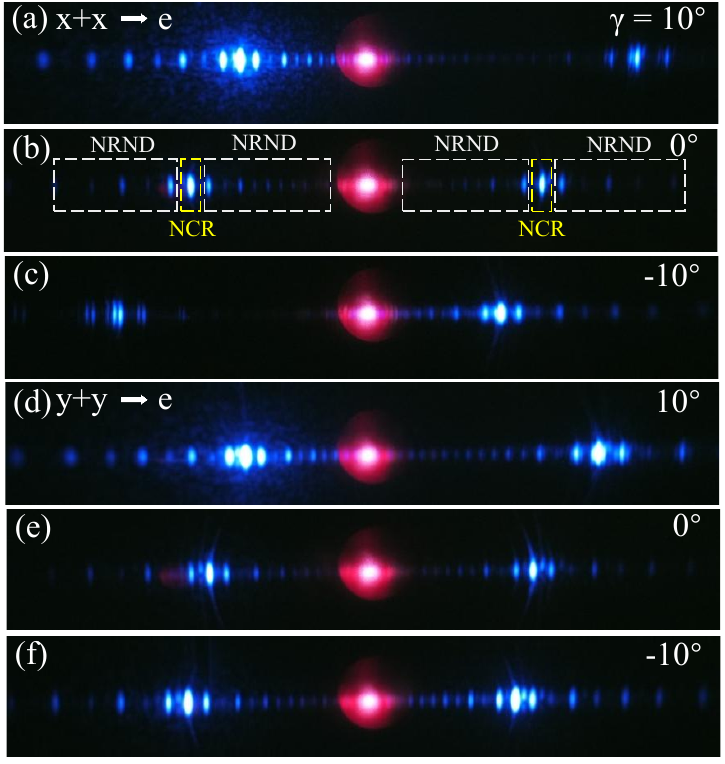}
\caption{NCR and $m^\mathrm{th}$‐order NRND patterns recorded in the PPKTP with a poling period of 9.83 $\mu$m at different polarizations and incident angles. (a)-(c) represent the second-harmonic diffraction patterns under the $x+x\to e$ with incident angles of -10°, 0°, and 10°; (d)-(e) represent the patterns under the $y+y\to e$ with incident angles of -10°, 0°, and 10°, respectively. The NCR and NRND are marked with boxes in (b) as an example. T=24°C and the pump power is 500 mW for all figures.
}
\label{fig::5}
\end{figure}

The experimental results are shown in Fig.\,\ref{fig::5}\text{(a-f)}, with incident angle from 10° to -10°.
Fig.\,\ref{fig::5}(b) shows the case of $x+x\to e$ configuration under  0° incident angle.
There are several spots present. The two brightest spots are NCR, while the other spots are NRND.
NCR is brighter than NRND, this is caused by the following reason: For NCR since $\Delta k_z$ =0 is fully satisfied in Eq.\,(\ref{eq55}), while for NRND, $\Delta k_z \neq 0 $ in Eq.\,(\ref{eq1010}) for NRND.
The distance between the two NCR spot is $72.0\pm0.7$ mm.
Fig.\,\ref{fig::5}(e)  shows the case of $y+y\to e$ configuration under  0° incident angle.
The distance between the two NCR spot is $66.0\pm0.7$ mm, which is smaller than the case of $x+x\to e$ configuration in Fig.\,\ref{fig::5}(b). Both $x+x\to e$ and $y+y\to e$ belongs to type-I phase matching condition.
Then, we rotate the incident angle from 10° to -10°. As shown in  Fig.\,\ref{fig::5} (a, c, d, f), it can be observed that NCR spots are shifting from left to right.

%
%
\begin{figure}[!thp]
\centering
\includegraphics[width= 0.75\textwidth]{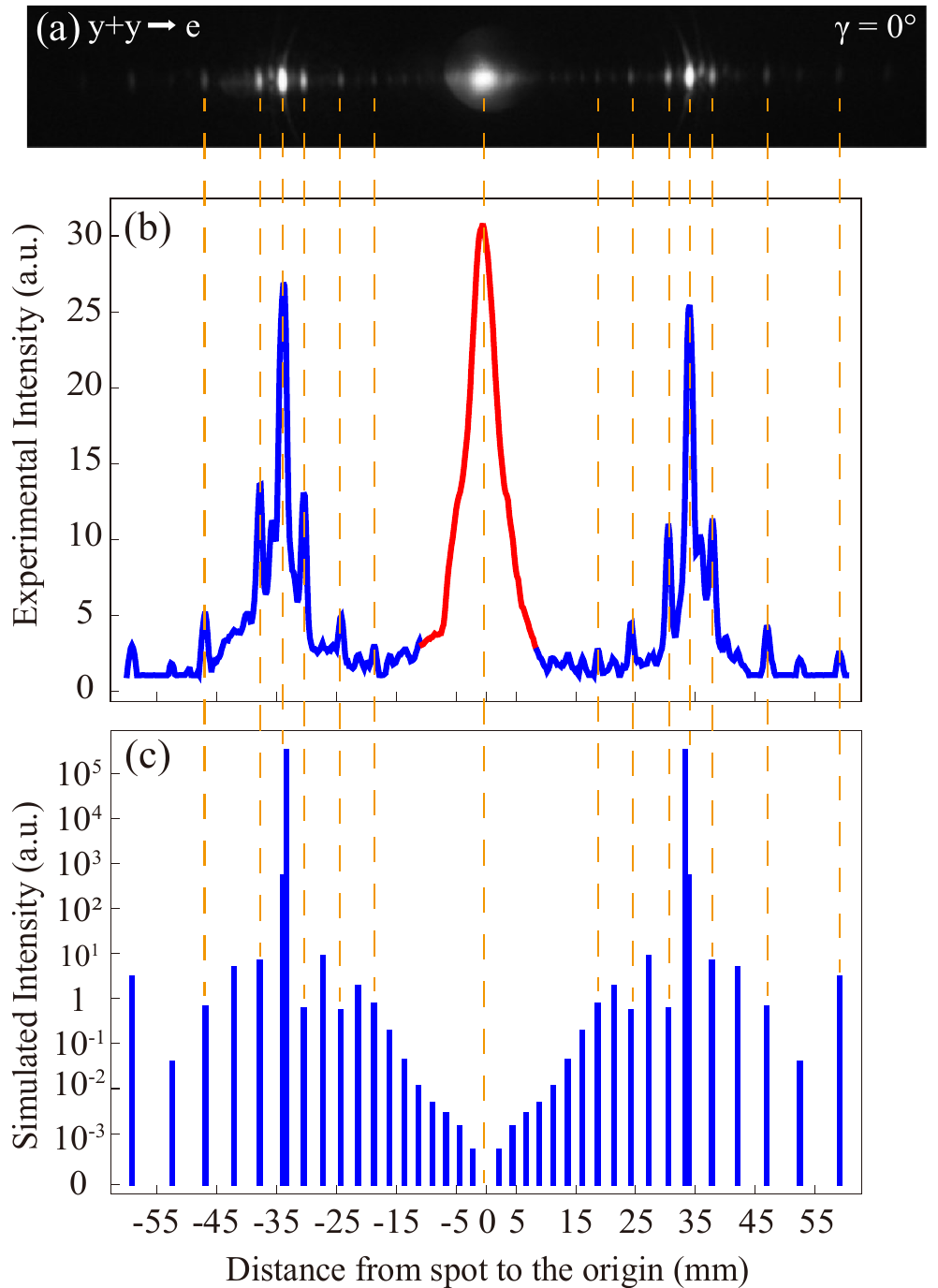}
\caption{Experimental and simulated nonlinear diffraction patterns for the $y+y\to e$ configuration at an incident angle of 0°. The crystal thickness $L_1$ is 1 mm, the distance from the crystal's surface B to the screen $L_2$ is 54.0 mm, and the poling period of the  PPKTP crystal is 9.83 $\mu$m. (a) The grayscale distribution of Fig.\,\ref{fig::5}(e); (b) the projection of (a) on the horizontal axis; (c) the theoretical simulation result of (b).
}
\label{Intnesity and position simulation}
\end{figure}
%
%

To verify the accuracy of our experiment, we also interpret the experimental results using theoretical calculations. Taking Fig.\,\ref{fig::5}(e) as an example, we first convert its color distribution into a grayscale distribution, as shown in Fig.\,\ref{Intnesity and position simulation}(a). Next, we project the two-dimensional distribution onto the horizontal axis to obtain a one-dimensional distribution, as shown in Fig.\,\ref{Intnesity and position simulation}(b). In this plot, the central red curve represents the unfiltered pump, while the blue curve corresponds to the NCR and NRND signals.
Next, we perform a theoretical simulation using the framework described in the Appendix. The calculated intensity distribution is shown in Fig.\,\ref{Intnesity and position simulation}(c). It can be seen that the positions of the experimental results correspond well with the theoretical simulations.
Notably, the NCR spot is positioned at $33.0\pm0.5$ mm from the center, which is in excellent accordance with the theoretical value of 33.9 mm. 
The experimental intensities diverge from the simulated outcomes, this might be caused by the reason that the duty cycle is perfectly 50:50.

%
%
\begin{figure}[!thp]
\centering
\includegraphics[width= 0.75\textwidth]{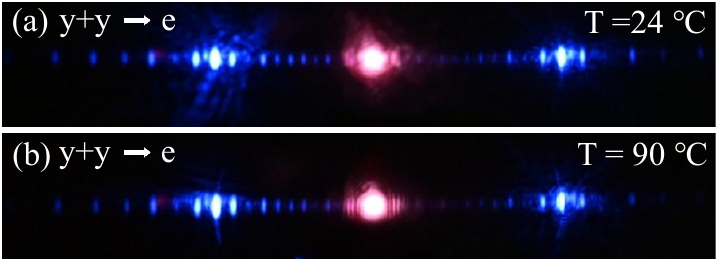}
\caption{NCR and NRND patterns recorded in the PPKTP for the $y+y\to e$ configuration with a poling period of 9.83 $\mu$m at different temperatures: (a) T=24°C and (b) T=90°C. The pump power is 500 mW.
}
\label{result temperature}
\end{figure}
%
%

Next, we investigate the temperature-dependent properties of NCR and NRND. The results are shown in Fig.\,\ref{result temperature}, where (a) corresponds to room temperature (24°C) and (b) corresponds to 90°C. It can be observed that the positions are almost unchanged with the increase in temperature. Therefore, we conclude that the NCR and NRND in the PPKTP crystal are insensitive to temperature.

%
\begin{figure}[!thp]
\centering
\includegraphics[width= 0.75\textwidth]{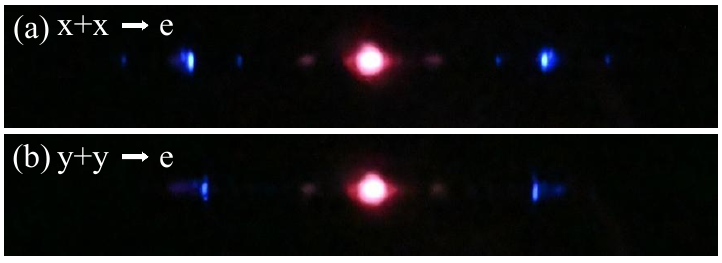}
\caption{
NCR and NRND patterns generated using a PPKTP crystal with a poling period of 3.43 $\mu$m at an  incident angle of 0°,  for two different polarizations: (a) $x+x\to e$ and (b) $y+y\to e$. T=24°C and the pump power is 500 mW.
}
\label{result poling peroid}
\end{figure}
%
%
%

To investigate the effect of the poling period of PPKTP crystals on nonlinear diffraction, we used a different PPKTP crystal with a poling period of 3.43 $\mu$m, while maintaining the same size of 10×2×1 mm. The crystal was pumped at 0° at room temperature, and the experimental results are shown in Fig.\,\ref{result poling peroid}. Figures (a) and (b) display the nonlinear diffraction patterns observed under x-polarization and y-polarization, respectively.
Furthermore, by comparing Fig.\,\ref{result poling peroid}(a) with Fig.\,\ref{fig::5}(b), and Fig.\,\ref{result poling peroid}(b) with Fig.\,\ref{fig::5}(e), we can see that the position of NCR remains almost unchanged for different poling periods, whereas the positions of NRND are substantially altered, because that longitudinal phase matching is invariant with respect to the poling period, and the transverse phase matching is dependent on this parameter. 

\begin{figure}[!thp]
\centering
\includegraphics[width= 0.75\textwidth]{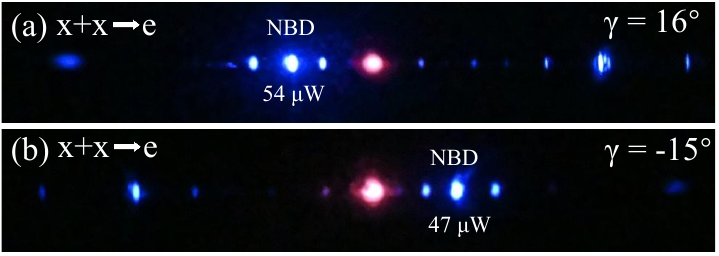}
\caption{The demonstration of nonlinear Bragg diffraction. Here the $x+x\to e$ configuration is utilized, with a PPKTP crystal having a poling period of 3.43 $\mu$m. The NBD phenomenon was observed at two different incident angles: (a) $+$16° and (b) $-$15°.  T=24°C and the pump power is 200 mW.}
\label{discussion NBD}
\end{figure}

\begin{figure}[!thp]
\centering
\includegraphics[width= 0.65\textwidth]{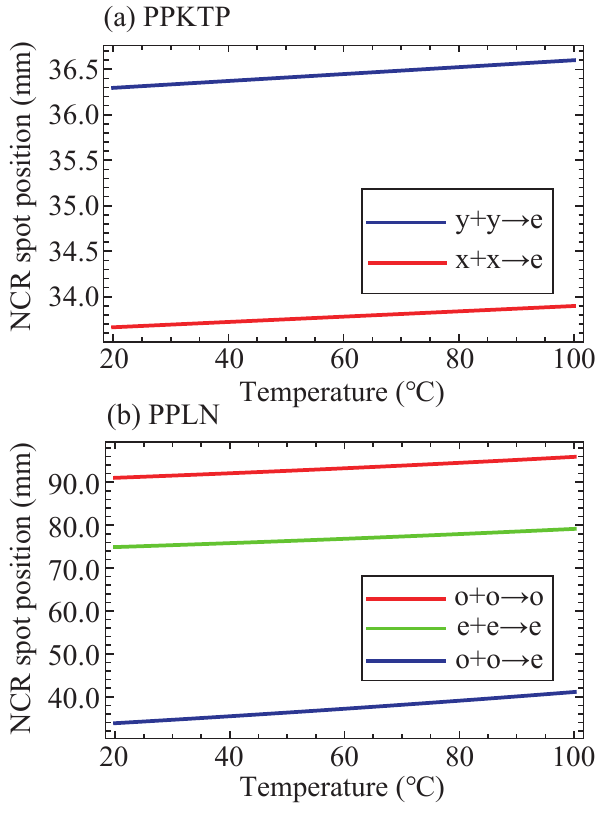}
\caption{NCR spot position at different temperature for PPKTP in (a) and for PPLN in (b). }
\label{discussion Temp}
\end{figure}

Another interesting phenomenon is NBD, which occurs when both transverse and longitudinal phase matching conditions are simultaneously met.
In our experiment, by rotating the crystal to change the pump laser's angle of incidence, we observed the NCR spot successively overlapping with the NRND spots of each order, which led to a sharp increase in light intensity. Fig.\,\ref{discussion NBD} shows the strongest NBD signal we observed, with corresponding powers of 54 $\mu$W and 47 $\mu$W, using a power meter instead of a camera. These signals were found at incident angles of 16°$\pm$ 0.5° and -15°$\pm$ 0.5°, respectively, which corresponds well with the theoretical calculation value $\pm$17.3°.

\section{Discussion}
First, we compare the thermal effects of PPKTP  and PPLN.
Fig.\,\ref{discussion Temp}(a) shows the calculated NCR spot position for PPKTP as the temperature varies from 20°C to 100°C.
In the $x+x\to e$ configuration, the NCR spot position shifts from 36.7 mm to 37.0 mm, corresponding to an average shift of 4 $\mu$m /°C.
Similarly, in the $y+y\to e$ configuration, the  average displacement is 3 $\mu$m /°C.
In contrast,  the PPLN crystal demonstrates large variations in temperature.
As shown in Fig.\,\ref{discussion Temp}(b), the average displacement are  60 $\mu$m /°C, 52 $\mu$m /°C, and 90 $\mu$m /°C for the configuration of  $o+o\to o$,  $e+e\to e$, and $o+o\to e$, receptively.
This implies that PPKTP crystals exhibit more than ten times greater thermal insensitivity compared to PPLN crystals. These results are also consistent with our experimental data presented in Fig.\,\ref{result temperature}.
In this simulation, we assumed that the $L_1=1$ mm and $L_2=54.0$ mm, as depicted in Fig.\,\ref{fig:theory}. The Sellmeier equation for PPLN is referenced from \cite{gayer2008APB}. 

Next, we discuss the advantages of biaxial crystals over uniaxial crystals in NCR and NRND processes. Uniaxial crystals possess a single optical axis and two distinct refractive indices, $n_o \neq n_e$. In contrast, biaxial crystals have two optical axes and three mutually different principal refractive indices, $n_x \neq n_y \neq n_z$.
As a consequence, a uniaxial crystal has only one principal plane and supports six phase-matching configurations ($oo-o, ee-e, oe-o, oe-e, oo-e, ee-o$) \cite{hong2024JOSAB}. By comparison, a biaxial crystal has three principal planes and allows eighteen phase-matching configurations: in the YOZ principal plane ($xx-e, xe-e, ee-e, xx-x, xe-x, ee-x$), in the XOZ principal plane ($yy-e, ye-e, ee-e, yy-y, ye-y, ee-y$), and in the XOY principal plane ($zz-e, ze-e, ee-e, zz-z, ze-z, ee-z$) \cite{smith2018crystal}.
Therefore, biaxial crystals provide significantly richer phase-matching possibilities and support a wider variety of frequency-conversion processes in NCR and NRND.

Finally, we consider potential future applications. Both simulation and experimental results demonstrate that the nonlinear diffraction of PPKTP can effectively fan out multiple optical channels. Furthermore, its spatial modes exhibit exceptionally high thermal stability, surpassing that of PPLN by more than an order of magnitude. This feature is critical for all-optical parallel network computing \cite{zhu2022NC, hu2024NC} and vortex light generation in domain engineering \cite{liu2019NC, yang2023OL}.
Conventional nonlinear frequency conversion devices typically rely on sophisticated and precise temperature-control systems to maintain modal stability. In contrast, PPKTP maintains highly stable spatial diffraction patterns without the need for elaborate thermal control, even when operated at high power or in elevated temperature conditions. Such remarkable robustness ensures accurate and stable coupling of the diffracted light into detector arrays, effectively suppressing optical path deviations caused by thermally induced beam deflection and significantly reducing the bit error rate of the optical computing system.

\section{Conclusion}
In conclusion, we conducted a comprehensive experimental examination of the distribution of NCR and NRND using PPKTP crystals. We utilized two different poling periods, 9.83 $\mu$m and 3.43 $\mu$m, to systematically analyze the effects of various parameters, including pump laser polarizations, incident angles, and crystal temperatures on both NCR and NRND.
Our results demonstrate that the polarization and incident angle of the pump laser significantly influence the positions of NCR and NRND. In contrast, while changes in the poling period do not affect the position of NCR, they do impact both the position and sequence of NRND.
More importantly, we found that PPKTP (3 $\mu$m /°C) exhibits over ten times greater thermal stability compared to PPLN (52 $\mu$m /°C). This high thermal stability holds promise for applications in parallel optical computing, as it helps reduce optical mode deviations and minimizes bit error rates.

\section*{Acknowledgments}
This work was supported by the National Natural Science Foundations of China (Grant Numbers  12574389 and 92365106).

\renewcommand\thefigure{A\arabic{figure}}
\setcounter{figure}{0}

\setcounter{equation}{0}
\renewcommand\theequation{A\arabic{equation}}

\clearpage

\appendix

\section{Theory} 
In this section, we present theoretical framework for the NRND and NCR in a PPKTP crystal. 

\label{app1}
\subsection{NCR and NRND  at $0^\circ$ incident angle}
During the SHG process, two pump photons with longer wavelength $\lambda_{1}$ enter a PPKTP crystal, resulting in the generation of a photon with a wavelength $\lambda_{2}$. This process can be characterized by: $\lambda_{1}$+$\lambda_{1}$$\to$$\lambda_{2}$, and in this work we set 810 nm+810 nm$\to$405 nm. From the viewpoint of polarization, two processes are possible: $x+x\to e$ for $x$ polarization, or $y+y\to e$ for $y$ polarization, here $e$ denotes the extraordinary-ray. 
The Sellmeier equation of PPKTP crystal can be found in reference \cite{kato2002KTPsellmeier}.
We start our analysis from the simplest case, i.e., incident angle is 0, and then expand the analysis to more complex case, i.e., incident angle is $\gamma$.
%
%
\begin{figure}[!thp]
\centering
\includegraphics[width= 0.65\textwidth]{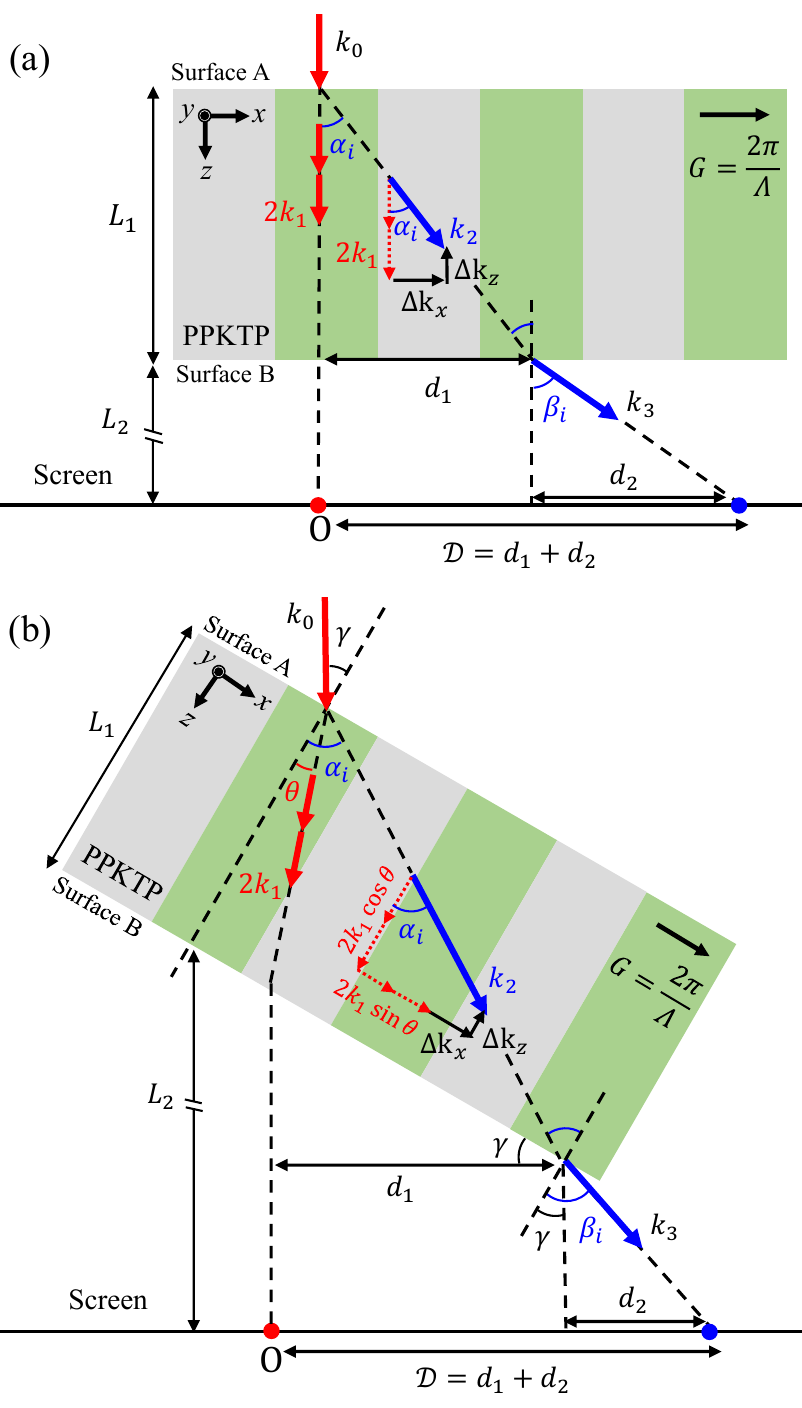}
\caption{ Schematic diagram of the pump laser passing through the PPKTP crystal. (a) incident angle of $0^\circ$, (b) incident angle of $\gamma$. In the figure, $\alpha_i$ and $\beta_i$ represent the nonlinear diffraction angles in the PPKTP crystal and in air, respectively, where $i=(\text{NCR, NRND})$.}
\label{fig:theory}
\end{figure}
%
%
%

%
%
%
The configuration of NCR and NRND is illustrated in  Fig.\,\ref{fig:theory}(a) for normal incidence. For the NCR, the longitudinal phase matching  can be written as:
\begin{equation}
\label{eq1}
\begin{split}
\Delta k_z=k_2\cos\alpha_{NCR}-2k_{1},
\end{split} 
\end{equation}
where $k_{2}$ is the wave vector of the second harmonic laser in PPKTP;  $k_{1}$ denotes the  wave vector of the pump laser in the PPKTP crystal; $\alpha_{NCR}$, the diffraction angle of NCR, refers to the angles formed between $k_{2}$ and the z-axis. 
When $\Delta k_z=0$, the diffraction angle can be calculated:
\begin{equation}
\label{eq2}
\begin{split}
\alpha_{NCR}=\arccos(\frac{2k_{1}}{k_{2}}).
\end{split} 
\end{equation}

For NRND, the transverse phase matching is:
\begin{equation}
\label{eq3}
\begin{split}
\Delta k_x=k_2\sin\alpha_{NRND}\pm mG,
\end{split} 
\end{equation}
where $m=0, \pm1, \pm2,\pm3 ...$ represents different  orders of NRND, $\alpha_{NRND}$ is the diffraction angle of the $m^\mathrm{th}$-order NRND; $G={2\pi}/{\Lambda}$ is the reciprocal lattice vector; $\Lambda$ is the poling period of the crystal.
When $\Delta k_x=0$, the diffraction angle of the $m^\mathrm{th}$-order NRND can be calculated as:
\begin{equation}
\label{eq4}
\begin{split}
\alpha_{NRND}=\arcsin{(\pm\frac{mG}{k_{2}})}.
\end{split} 
\end{equation}

\subsection{NCR and NRND  at $\gamma$ incident angle}
Then, we rotate the crystal by an angle of $\gamma$, as  illustrated in  Fig.\,\ref{fig:theory}(b). Under this condition, the longitudinal phase matching for NCR is:
\begin{equation}
\label{eq55}
\begin{split}
\Delta k_z=k_2\cos\alpha_{NCR}-2k_{1}\cos\theta,
\end{split} 
\end{equation}
where $\theta$ denotes the angle formed between  $k_{1}$ and the z-axis. $\theta$ can be derived using the law of Snell's law:
\begin{equation}
\label{eq6}
\begin{split}
  \theta=\arcsin(\frac{\sin\gamma}{n_1}),
\end{split} 
\end{equation}
where, $n_{1}$ represents the refractive index of the pump laser in the PPKTP.

When  $\Delta k_z=0$, the diffraction angle $\alpha_{NCR}$ can be calculated as follow:
\begin{equation}
\label{eq77}
\begin{split}
\alpha_{NCR}=\arccos(\frac{2k_{1}\cos{\theta}}{k_{2}}).
\end{split} 
\end{equation}

Similarly, we can calculate the transverse phase matching condition $\Delta k_x$: 
\begin{equation}
\label{eq7}
\begin{split}
\Delta k_x=k_2\sin\alpha_{NRND}-2k_{1}\sin\theta\pm mG,
\end{split} 
\end{equation}
where $\alpha_{NRND}$  is the diffraction angle of $m^\mathrm{th}$-order NRND, and can be calculated as follow
\begin{equation}
\label{eq8}
\begin{split}
\alpha_{NRND}=\arcsin(\frac{2k_{1}\sin\theta\pm {m G}}{k_{2}}).
\end{split} 
\end{equation}
Similarly, the longitudinal phase matching for NRND is:
\begin{equation}
\label{eq1010}
\begin{split}
\Delta k_z=k_2\cos\alpha_{NRND}-2k_{1}\cos\theta,
\end{split} 
\end{equation}

\subsection{The spatial position of NCR and NRND}
In the previous section, the diffraction angles associated with NCR and NRND within the crystal were derived. Now, we calculate the position of the nonlinear diffraction spots on the screen, as shown in Fig.\,\ref{fig:theory}(b).

The second harmonic  is refracted on surface B, satisfying:
\begin{equation}
\label{eq10}
\begin{split}
  \beta_{i}=\arcsin(n_{2}\sin\alpha_{i}),
\end{split} 
\end{equation}
where $n_2$ is the refractive index of the $k_2$ in the PPKTP crystal, which can be calculated as follow:
\begin{equation}
\label{eq11}
\begin{split}
  n_{2} = \sqrt{\frac{1}{\frac{\cos^2\alpha_i}{n_x^2} + \frac{\sin^2\alpha_i}{n_z^2}}}.
\end{split} 
\end{equation}
Note, in this work, we focus on the xoz plane as shown in Fig.\,\ref{fig:d_dff}.

The projection point of the incident laser on the screen is the origin ($\mathcal{D}=0$), the crystal thickness is $L_{1}$, and the distance between the crystal surface B and the screen is $L_2$.
The parameter $\mathcal{D}$ denotes the distance from each diffraction spot to the origin on the projection screen. 
\begin{equation}
\label{eq13x}
   \mathcal{D}=d_1+d_2.
\end{equation}
Here, $d_{1}$ is the lateral offset distance of $k_{2}$ within the crystal:
\begin{equation}
\label{eq13}
   d_{1}=L_{1}\times(\tan\alpha_i-\tan\theta)\times\cos\gamma,
\end{equation}
and  $d_{2}$ indicates the lateral offset distance of $k_{3}$ within the air:
\begin{equation}
\label{eq14}
    d_{2}=(L_2-L_1\times\tan\alpha_i\times\sin\gamma)
    \times\tan(\beta_i-\gamma).
\end{equation}
\subsection{The intensity of NCR and NRND}
The intensity of the second harmonic laser of the nonlinear diffraction spots is $I$ \cite{hong2024JOSAB}: 
\begin{align}
\label{eq15}
\begin{split}
I=\left[\frac{\omega_2^2 \times\chi_{eff}^{(2)}\times E_1^2\times L_{1}}{2\times\mathrm{k}_2 \times\cos\alpha_i\times c^2}\times \mathrm{sinc}(\frac{\Delta k_z\times L_{1}}{2})\right]^2,
\end{split} 
\end{align}
where $\omega_{2}$ the angular frequency of $k_{2}$, and  $\omega_{2}=\frac{2\pi c}{\lambda_{2}}$. $\alpha_i$ represents the diffraction angles of nonlinear diffraction; $c$ is the speed of light; $E_{1}$ is the amplitude of the pump laser; $\chi_{eff}^{(2)}$ is the second-order nonlinear susceptibility. 
From the equation above, we can notice that the intensity of NRND and NCR is directly determined by $\Delta k_z$, not $\Delta k_x$.
$\chi_{eff}^{(2)}$ can be expressed by the effective nonlinear coefficient $d_{eff}$ as follow \cite{hong2024JOSAB}:

\begin{equation}
\label{eq16}
\begin{array}{ll}
   \chi_{eff}^{(2)}=2\times d_{eff}.
\end{array}
\end{equation}

For PPKTP crystal, the second order nonlinear coefficient tensor $d$ can be written \cite{Pack2004AO}: 
\begin{equation}
\begin{aligned}
d&=
\begin{bmatrix}
0 & 0 & 0 & 0 & d_{xxz} & 0 \\
0 & 0 & 0 & d_{yyz} & 0 & 0 \\
d_{zxx} & d_{zyy} & d_{zzz} & 0 & 0 & 0
\end{bmatrix}
\\&=
\begin{bmatrix}
0 & 0 & 0 & 0 & 1.95\;\, & 0 \\
0 & 0 & 0 & 3.9\;\, & 0 & 0 \\
1.95 \;\, & 3.9\;\, & 15.3 \;\,& 0 & 0 & 0
\end{bmatrix}.
\end{aligned}
\end{equation}

%
%
\begin{figure}[!thp]
\centering
\includegraphics[width= 0.65\textwidth]{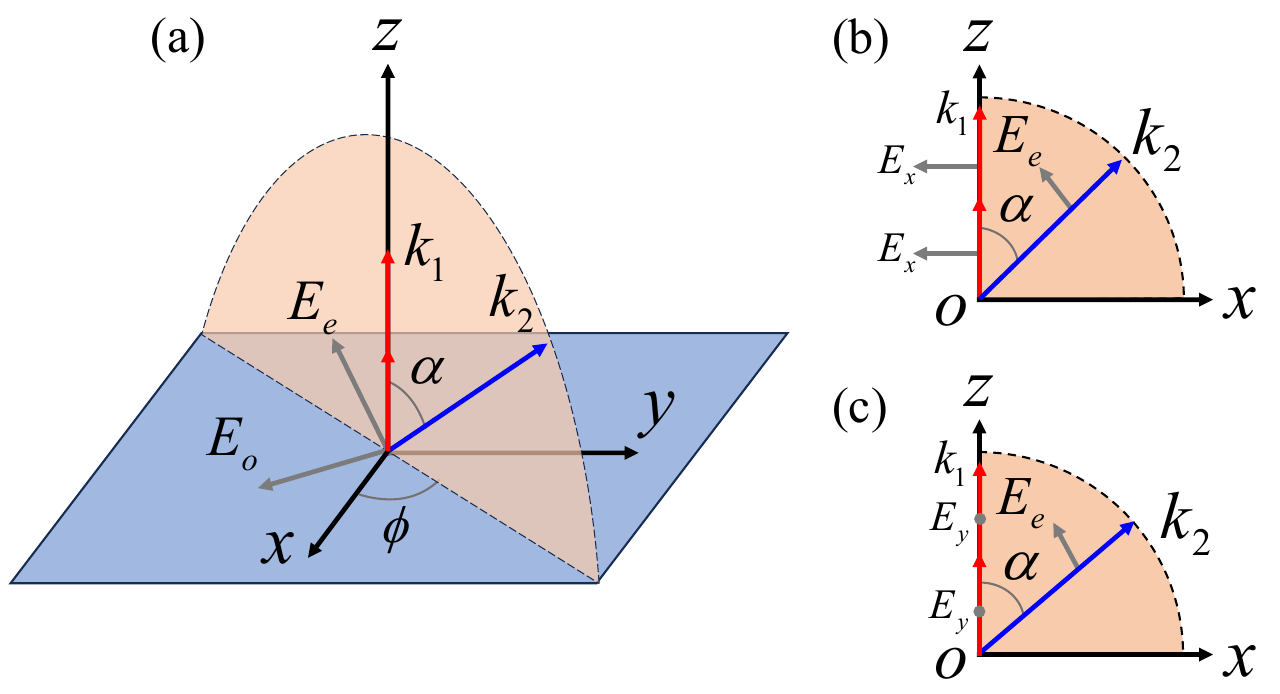}
\caption{(a) Vector distribution in NCR and NRND; (b) vector distribution of $x+x\to e$; (c) vector distribution of $y+y\to e$.
}
\label{fig:d_dff}
\end{figure}
%
%

In nonlinear diffraction, $d_{eff}$ is influenced by both the diffraction angle $\alpha$ and the polarizations of the pump and second-harmonic fields, as shown in Fig.\,\ref{fig:d_dff}.
In the case of $x+x\to e$ configuration, as shown in Fig.\,\ref{fig:d_dff}(b), the $d_{eff}$  can be expressed as \cite{smith2018crystal}:
\begin{equation}
\label{eq18}
\begin{array}{ll}
d_{eff} &= -d_{xxx}\cos^3\alpha - d_{xzz}\cos\alpha\sin^2\alpha \\
        &\phantom{=}+ 2d_{xxz}\cos^2\alpha\sin\alpha
          + d_{zxx}\cos^2\alpha\sin\alpha \\
        &\phantom{=}+ d_{zzz}\sin^3\alpha 
          - 2d_{zxz}\cos\alpha\sin^2\alpha \\
        &= 2d_{xxz}\cos^2\alpha\sin\alpha
          + d_{zxx}\cos^2\alpha\sin\alpha \\
        &\phantom{=}+ d_{zzz}\sin^3\alpha.
\end{array}
\end{equation}
Note here $\alpha$ can be $\alpha_{NCR}$ or $\alpha_{NRND}$.

The $d_{eff}$ for the $y+y \to e$, as shown in Fig.\,\ref{fig:d_dff}(c), can be written as \cite{smith2018crystal,hong2024JOSAB}:
\begin{equation}
\label{eq19}
\begin{array}{ll}
   d_{eff}&=-d_{xyy}\cos\alpha+d_{zyy}\sin\alpha\\&=d_{zyy}\sin\alpha,
\end{array}
\end{equation}

From equations (\ref{eq18}) and (\ref{eq19}), it can be seen that the $d_{eff}$ is related to the diffraction angle $\alpha$. Taking a diffraction angle of 30° as an example: for the $x+x\to e$ case, $d_{eff}=3.37 \, {pm/V}$ and for the $y+y\to e$ case, $d_{eff}=1.95 \, {pm/V}$. In addition, we found that in a PPKTP crystal, when the pump laser propagates along the z-axis, the effective nonlinear coefficient is zero for both the $x+x\to o$ and $y+y\to o$ cases, here o denotes the ordinary-ray.

\section{Spectrum} 
\begin{figure}[!thp]
\centering
\includegraphics[width= 0.98\textwidth]{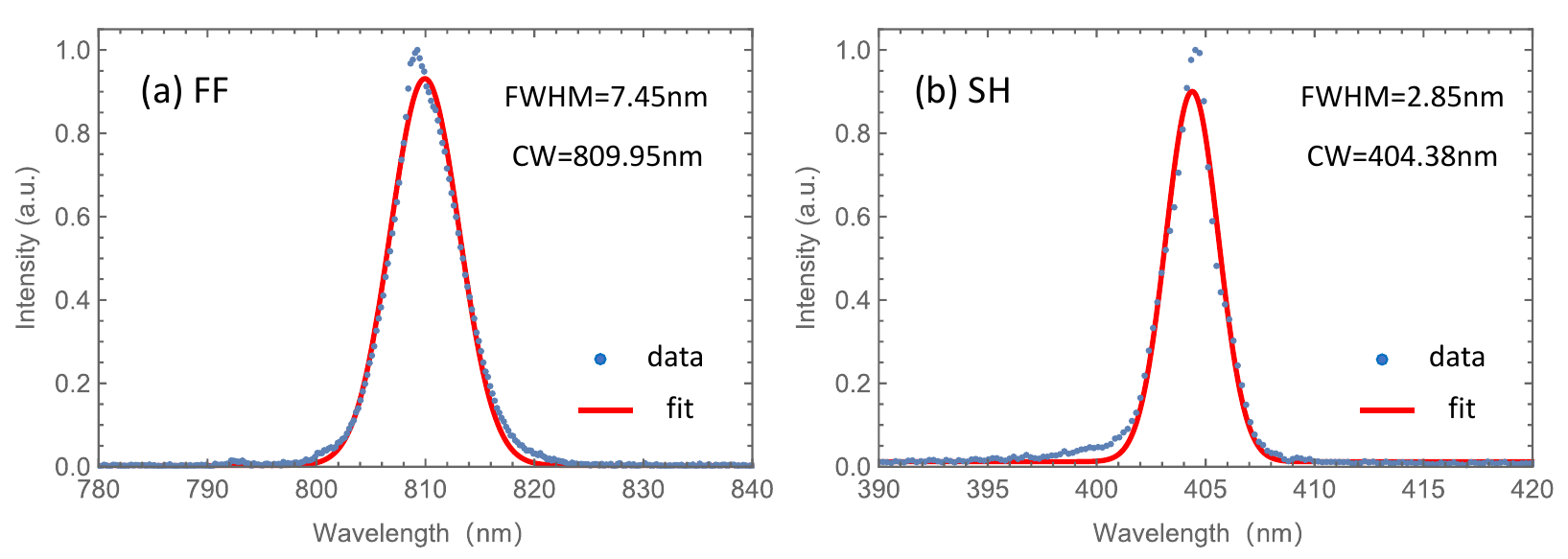}
\caption{(a) Spectrum of the fundamental frequency light and (b) spectrum of the second harmonic light. Experimental data are represented by blue dots, while the red solid lines indicate the corresponding numerical fits. CW: center wavelength; FWHM: full width at half maximum.
}
\label{fig:Spectrum}
\end{figure}

The experimental spectra of the fundamental frequency (FF) and second harmonic (SH) light are displayed in Fig.\,\ref{fig:Spectrum}(a) and (b), respectively. These measurements were conducted using a spectrometer (BIM-6001-06, Brolight Co.), which features a spectral range of 350–1050 nm and a typical resolution of 0.8 nm.

%
%
%
%

\bibliographystyle{elsarticle-num-names} 






\bibliography{2023LatexTemplate}

@article{vyunishev2014OL,
  title={{Nonlinear Raman--Nath diffraction of femtosecond laser pulses}},
  author={Vyunishev, AM and Slabko, VV and Baturin, IS and Akhmatkhanov, AR and Shur, V Ya},
  journal={Optics Letters},
  volume={39},
  number={14},
  pages={4231--4234},
  year={2014},
  publisher={Optical Society of America}
}

@article{Jin2024SCPMA,
  title={Spectrally resolved Franson interference},
  author={Rui-Bo Jin and Zi-Qi Zeng and Dan Xu
and Chen-Zhi Yuan and Bai-Hong Li and You Wang
and Ryosuke Shimizu and Masahiro Takeoka and Mikio Fujiwara and Masahide Sasaki and Pei-Xiang Lu},
  journal={Sci. China: Phys., Mech. Astron.},
  volume={67},
  number={5},
  pages={250312},
  year={2024},
  publisher = {Springer {US}}
}

@article{Zeng2024COL,
  title={Controllable transitions among phase-matching conditions in a single nonlinear crystal},
  author={Zeng, Ziqi and You, Shixin and Yang, Zixiang and Yuan, Chenzhi and You, Chenglong and Jin, Ruibo},
  journal={Chinese Optics Letters},
  volume={22},
  number={2},
  pages={021901},
  year={2024}
}

@article{li2024PhotoniX,
  title = {Nonlinear Raman-Nath Diffraction in Submicron-Thick Periodically Poled Lithium Niobate Thin Film},
  author = {Li, Xiao-Ni and Peng, Ling-Zhi and Liu, Yuan-Yuan and Hong, Li-Hong and Hu, De-Ming and Zhao, Yuan-Yuan and Duan, Xuan-Ming and Chen, Bao-Qin and Li, Zhi-Yuan},
  year = {2024},
  journal = {Photonix},
  volume = {5},
  number = {1},
  pages = {42}
}

@article{shapira2011OL,
  title={Phase-matched nonlinear diffraction},
  author={Shapira, Asia and Arie, Ady},
  journal={Optics Letters},
  volume={36},
  number={10},
  pages={1933--1935},
  year={2011},
  publisher={Optical Society of America}
}

@article{sheng2010OL,
  title={Cerenkov-type second-harmonic generation with fundamental beams of different polarizations},
  author={Sheng, Yan and Saltiel, Solomon M and Krolikowski, Wieslaw and Arie, Ady and Koynov, Kaloian and Kivshar, Yuri S},
  journal={Optics Letters},
  volume={35},
  number={9},
  pages={1317--1319},
  year={2010},
  publisher={Optical Society of America}
}

@article{saltiel2009IEEE,
  title={Cerenkov-type second-harmonic generation in two-dimensional nonlinear photonic structures},
  author={Saltiel, Solomon Mois and Sheng, Yan and Voloch-Bloch, Noa and Neshev, Dragomir N and Krolikowski, Wieslaw and Arie, Ady and Koynov, Kaloian and Kivshar, Yuri S},
  journal={IEEE Journal of Quantum Electronics},
  volume={45},
  number={11},
  pages={1465--1472},
  year={2009},
  publisher={IEEE}
}

@article{saltiel2009OL,
  title={Multiorder nonlinear diffraction in frequency doubling processes},
  author={Saltiel, Solomon M and Neshev, Dragomir N and Krolikowski, Wieslaw and Arie, Ady and Bang, Ole and Kivshar, Yuri S},
  journal={Optics Letters},
  volume={34},
  number={6},
  pages={848--850},
  year={2009},
  publisher={Optical Society of America}
}

@article{holmgren2007OL,
  title={Ultrashort single-shot pulse characterization with high spatial resolution using localized nonlinearities in ferroelectric domain walls},
  author={Holmgren, SJ and Canalias, Carlota and Pasiskevicius, Valdas},
  journal={Optics Letters},
  volume={32},
  number={11},
  pages={1545--1547},
  year={2007},
  publisher={Optical Society of America}
}

@article{aleksandrovsky2011APL,
  title={Ultrashort pulses characterization by nonlinear diffraction from virtual beam},
  author={Aleksandrovsky, AS and Vyunishev, AM and Zaitsev, AI and Ikonnikov, AA and Pospelov, GI},
  journal={Applied Physics Letters},
  volume={98},
  number={6},
  year={2011},
  publisher={AIP Publishing}
}

@article{ren2013APL,
  title={Surface phase-matched harmonic enhancement in a bulk anomalous dispersion medium},
  author={Ren, Huaijin and Deng, Xuewei and Zheng, Yuanlin and An, Ning and Chen, Xianfeng},
  journal={Applied Physics Letters},
  volume={103},
  number={2},
  year={2013},
  publisher={AIP Publishing}
}

@article{karpinski2015OE,
  title={Nonlinear diffraction in orientation-patterned semiconductors},
  author={Karpinski, Pawel and Chen, Xin and Shvedov, Vladlen and Hnatovsky, Cyril and Grisard, Arnaud and Lallier, Eric and Luther-Davies, Barry and Krolikowski, Wieslaw and Sheng, Yan},
  journal={Optics Express},
  volume={23},
  number={11},
  pages={14903--14912},
  year={2015},
  publisher={Optical Society of America}
}

@article{deng2010OE,
  title={Domain wall characterization in ferroelectrics by using localized nonlinearities},
  author={Deng, Xuewei and Chen, Xianfeng},
  journal={Optics Express},
  volume={18},
  number={15},
  pages={15597--15602},
  year={2010},
  publisher={Optical Society of America}
}

@article{kalinowski2012OL,
  title={{Enhanced Cerenkov second-harmonic emission in nonlinear photonic structures}},
  author={Kalinowski, Ksawery and Roedig, P and Sheng, Yan and Ayoub, M and Imbrock, Jorg and Denz, Cornelia and Krolikowski, Wieslaw},
  journal={Optics Letters},
  volume={37},
  number={11},
  pages={1832--1834},
  year={2012},
  publisher={Optical Society of America}
}

@article{hong2022PhotonicsResearch,
  title={{Ultrabroadband nonlinear Raman--Nath diffraction against femtosecond pulse laser}},
  author={Hong, Lihong and Chen, Baoqin and Hu, Chenyang and Li, Zhi-Yuan},
  journal={Photonics Research},
  volume={10},
  number={4},
  pages={905--912},
  year={2022},
  publisher={Chinese Laser Press and Optica Publishing Group}
}

@article{sheng2012JOSAB1,
  title = {Theoretical Study of {Cerenkov}-Type Second-Harmonic Generation in Periodically Poled Ferroelectric Crystals},
  author = {Sheng, Yan and Kong, Qian and Roppo, Vito and Kalinowski, Ksawery and Wang, Qi and Cojocaru, Crina and Krolikowski, Wieslaw},
  year = {2012},
  journal = {Journal of the Optical Society of America B: Optical Physics},
  volume = {29},
  number = {3},
  pages = {312--318}
}

@article{zhao2016OE,
  title={{Nonlinear Cherenkov radiation at the interface of two different nonlinear media}},
  author={Zhao, Xiaohui and Zheng, Yuanlin and Ren, Huaijin and An, Ning and Deng, Xuewei and Chen, Xianfeng},
  journal={Optics Express},
  volume={24},
  number={12},
  pages={12825--12830},
  year={2016},
  publisher={Optical Society of America}
}

@article{sheng2012JOSAB2,
  title = {Theoretical Investigations of Nonlinear {Raman--Nath} Diffraction in the Frequency Doubling Process},
  author = {Sheng, Yan and Kong, Qian and Wang, Wenjie and Kalinowski, Ksawery and Krolikowski, Wieslaw},
  year = {2012},
  journal = {Journal of Physics B: Atomic, Molecular and Optical Physics},
  volume = {45},
  number = {5},
  pages = {55401}
}

@article{hong2022PRA,
  title={{Rainbow Cherenkov second-harmonic radiation}},
  author={Hong, Lihong and Chen, Baoqin and Hu, Chenyang and He, Peng and Li, Zhi-Yuan},
  journal={Physical Review Applied},
  volume={18},
  number={4},
  pages={044063},
  year={2022},
  publisher={APS}
}

@article{hong2025UScience,
  title={1.3-Octave Nonlinear {Cherenkov} Radiation Triggered by Intense 2.2-Octave Femtosecond White Laser},
  author={Hong, Lihong and Zou, Yu and Liu, Yuanyuan and Liu, Junming and Liu, Liqiang and Li, Zhi-Yuan},
  journal={Ultrafast Science},
  volume={5},
  pages={0102},
  year={2025},
  publisher={AAAS}
}

@article{zou2025COL,
  title={{Conical Raman--Nath nonlinear optical diffraction upon PPLN nonlinear gratings}},
  author={Zou, Yu and Hong, Lihong and Li, Jiacheng and Chen, Jianluo and Li, Zhi-Yuan},
  journal={Chinese Optics Letters},
  volume={23},
  number={7},
  pages={071902},
  year={2025}
}

@article{fragemann2004APL,
  title={{Second-order nonlinearities in the domain walls of periodically poled KTiOPO$_4$}},
  author={Fragemann, Anna and Pasiskevicius, Valdas and Laurell, Fredrik},
  journal={Applied Physics Letters},
  volume={85},
  number={3},
  pages={375--377},
  year={2004},
  publisher={AIP Publishing}
}

@article{an2013APL,
  title={Conical second harmonic generation in one-dimension nonlinear photonic crystal},
  author={An, Ning and Zheng, Yuanlin and Ren, Huaijin and Deng, Xuewei and Chen, Xianfeng},
  journal={Applied Physics Letters},
  volume={102},
  number={20},
  year={2013},
  publisher={AIP Publishing}
}

@article{zhang2021Optica,
  title={Nonlinear photonic crystals: from 2D to 3D},
  author={Zhang, Yong and Sheng, Yan and Zhu, Shining and Xiao, Min and Krolikowski, Wieslaw},
  journal={Optica},
  volume={8},
  number={3},
  pages={372--381},
  year={2021},
  publisher={Optical Society of America}
}

@article{zhang2018OE,
  title={Universal modeling of second-order nonlinear frequency conversion in three-dimensional nonlinear photonic crystals},
  author={Zhang, Jing and Zhao, Xiaohui and Zheng, Yuanlin and Li, Honggen and Chen, Xianfeng},
  journal={Optics Express},
  volume={26},
  number={12},
  pages={15675--15682},
  year={2018},
  publisher={Optical Society of America}
}

@article{yang2017OE,
  title={{Linear Cherenkov radiation in ferroelectric domain walls}},
  author={Yang, Ji and Zhao, Xiaohui and Liu, Haigang and Chen, Xianfeng},
  journal={Optics Express},
  volume={25},
  number={22},
  pages={27818--27823},
  year={2017},
  publisher={Optical Society of America}
}

@article{hong2024JOSAB,
  title={Unified analytical theory of nonlinear optical diffraction by nonlinear gratings},
  author={Hong, Lihong and Zou, Yu and Li, Jiacheng and Chen, Jianluo and Li, Zhi-Yuan},
  journal={Journal of the Optical Society of America B},
  volume={41},
  number={11},
  pages={2562--2588},
  year={2024},
  publisher={Optica Publishing Group}
}

@article{zhang2025COL,
  title={{Nonlinear Cherenkov radiation in rotatory nonlinear optics}},
  author={Zhang, Zhongmian and Lu, Dazhi and Yu, Haohai and Zhang, Huaijin and Wu, Yicheng},
  journal={Chinese Optics Letters},
  volume={23},
  number={4},
  pages={041901},
  year={2025}
}

@article{buono2022OEA,
  title={Nonlinear optics with structured light},
  author={Buono, Wagner Tavares and Forbes, Andrew},
  journal={Opto-Electronic Advances},
  volume={5},
  number={6},
  pages={210174--1},
  year={2022},
  publisher={Opto-Electronic Advances}
}

@article{gayer2008APB,
  title = {Temperature and Wavelength Dependent Refractive Index Equations for {MgO}-doped Congruent and Stoichiometric {LiNbO$_3$}},
  author = {Gayer, O. and Sacks, Z. and Galun, E. and Arie, A.},
  year = {2008},
  journal = {Applied Physics B},
  volume = {91},
  number = {2},
  pages = {343--348},
  publisher = {{Springer Science and Business Media LLC}}
}

@article{Pack2004AO,
  title = {Measurement of the {{$\chi$}}(2) Tensors of {KTiOPO$_4$}, {KTiOAsO$_4$}, {RbTiOPO$_4$}, and {RbTiOAsO$_4$} Crystals},
  author = {Pack, Michael V. and Armstrong, Darrell J. and Smith, Arlee V.},
  year = {2004},
  journal = {Applied Optics},
  volume = {43},
  number = {16},
  pages = {3319--3323},
  publisher = {Optica Publishing Group}
}

@book{smith2018crystal,
  title={Crystal nonlinear optics: with {SNLO} examples},
  author={Smith, Arlee V},
  year={2018},
  publisher={AS-Photonics Albuquerque, NM, USA}
}

@article{kato2002KTPsellmeier,
  title={Sellmeier and thermo-optic dispersion formulas for {KTP}},
  author={Kato, Kiyoshi and Takaoka, Eiko},
  journal={Applied Optics},
  volume={41},
  number={24},
  pages={5040--5044},
  year={2002},
  publisher={Optical Society of America}
}

@book{dmitriev2013handbook,
  title={{Handbook of nonlinear optical crystals}},
  author={Dmitriev, Valentin G and Gurzadyan, Gagik G and Nikogosyan, David N},
  volume={64},
  year={2013},
  publisher={Springer}
}

@article{niu2023COL,
  title = {Multi-Color Laser Generation in Periodically Poled {KTP} Crystal with Single Period},
  author = {Niu, Sujian and Zhou, Zhiyuan and Cheng, Jingxin and Ge, Zheng and Yang, Chen and Shi, Baosen},
  year = {2023},
  journal = {Chinese Optics Letters},
  volume = {21},
  number = {2},
  pages = {21901}
}

@article{zhu2022NC,
  title = {Space-Efficient Optical Computing with an Integrated Chip Diffractive Neural Network},
  author = {Zhu, H. H. and Zou, J. and Zhang, H. and Shi, Y. Z. and Luo, S. B. and Wang, N. and Cai, H. and Wan, L. X. and Wang, B. and Jiang, X. D. and Thompson, J. and Luo, X. S. and Zhou, X. H. and Xiao, L. M. and Huang, W. and Patrick, L. and Gu, M. and Kwek, L. C. and Liu, A. Q.},
  year = 2022,
  journal = {Nature Communications},
  volume = {13},
  number = {1},
  pages = {1044}
}

@article{hu2024NC,
  title = {Diffractive Optical Computing in Free Space},
  author = {Hu, Jingtian and Mengu, Deniz and Tzarouchis, Dimitrios C. and Edwards, Brian and Engheta, Nader and Ozcan, Aydogan},
  year = 2024,
  journal = {Nature Communications},
  volume = {15},
  number = {1},
  pages = {1525}
}

@article{liu2019NC,
  title = {Nonlinear Wavefront Shaping with Optically Induced Three-Dimensional Nonlinear Photonic Crystals},
  author = {Liu, Shan and Switkowski, Krzysztof and Xu, Chenglong and Tian, Jie and Wang, Bingxia and Lu, Peixiang and Krolikowski, Wieslaw and Sheng, Yan},
  year = 2019,
  journal = {Nature Communications},
  volume = {10},
  number = {1},
  pages = {3208},
}

@article{yang2023OL,
  title = {Highly Efficient Nonlinear Vortex Beam Generation by Using a Compact Nonlinear Fork Grating},
  author = {Yang, Yangfeifei and Li, Hao and Liu, Haigang and Chen, Xianfeng},
  year = 2023,
  journal = {Optics Letters},
  volume = {48},
  number = {24},
  pages = {6376},
}
\end{document}